\documentstyle[12pt,aps,prc,preprint,epsfig]{revtex}

\tightenlines

\begin{document}

\title{Applications of semiempirical shell model masses based on\\ magic
    proton number  Z = 126 to superheavy nuclei}
\author{S. Liran\footnote{Present address: Kashtan 3/3, Haifa 34984,
 Israel.}, A. Marinov and N. Zeldes}
%\date{}
\maketitle

\centerline{\it The Racah Institute of Physics, The Hebrew
University of  Jerusalem,}
 \centerline{\it Jerusalem 91904,
Israel}
% \centerline{$^{2}${\it }}
\bigskip

\begin{abstract}
    A recently published highly extrapolatable semiempirical shell model
    mass equation is shown to describe rather well the energies of
    several seemingly well identified
    $\alpha$-decay chains with known end product nuclei observed in
superheavy elements research. The
    equation is also applied to the interpretation problem of
    some recent hot fusion-evaporation experiments with unknown end
    products and several
    conceivable reaction channels. Some plausible interpretations are
    indicated.
\end{abstract}


\section{Introduction}

Recently \cite{lmz00,lmz01} a semiempirical shell model mass
equation (SSME) based on proton magic number Z = 126 \cite{lir73}
was shown to have a high predictive power in the interior of major
shell regions beyond lead\footnote{There are two such regions,
with $82 \leq$ N $\leq 126$ and with 126 $\leq$ N $\leq$ 184. We
refer to them respectively as region A and region B, like in refs.
[1,2].}. The equation has been proposed as a substitute inside the
above regions for the previous SSME \cite{liz76} which is based on
Z = 114 as the highest proton magic number and it stops there.
Tabulated masses with separation and decay energies for both
regions are available \cite{lmzla}.

In the present work the predicted masses (eq. (1) of ref.
\cite{lmz00}, referred to as eq. (1) in the sequel, and table B of
ref. \cite{lmzla}) are applied to the results of recent superheavy
elements (SHE) experiments. We first address the smoothness of the
predictions and their overall agreement with $Q_{\alpha}$
systematics. Then we test the agreement of the equation with the
data observed in several recent fusion-evaporation SHE experiments
where the produced nuclei are identified by their decay ties to
known daughter isotopes. Finally we address the interpretation
problem in some recent hot fusion experiments where the observed
decay chains do not connect the parent to a known daughter, and
several reaction channels are in principle possible.

\section{Oversmoothness of the equation inside region B}

Inside a shell region the mass
surface predicted by eq.~(1) is smoother than the empirical surface and
does not account for fine structure effects.

This is illustrated in fig. 1 showing Q$_\alpha$ systematics for
the heaviest N $\geq$ 140 even-Z nuclei from Pu through Z = 110
\cite{auw95}. Respective full and empty circles denote
experimental values and values estimated from systematics. The
small circles connected by thin lines show the predictions of
eq.~(1).

%\begin{figure}[h]
%\begin{center}
%\leavevmode \epsfxsize=10.5cm \epsfbox{zead3a.ps}
%\end{center} \caption{Q$_\alpha$ systematics of even-Z elements from Pu
%through Z = 110 for N $\geq$ 140. Data taken from ref. [6] and
%predictions from refs. [1,5].}
%\end{figure}

As a rule both the experimental and the predicted isotopic lines
show similar negative trends when N increases, and they shift
upwards rather uniformly when Z increases. For the experimental
lines, though, this regular pattern breaks down for nuclei in the
vicinity of the deformed doubly submagic nucleus (N$_{0}$,
Z$_{0}$) $^{252}$Fm (N$_{0}$ = 152, Z$_{0}$ = 100), and presumably
even more so near  $^{270}$Hs (N$_{0}$
 = 162, Z$_{0}$ = 108) \cite{mon94,sob96,smo97,laz96} (see also fig. 5.). In these
 neighborhoods  the trend of isotopic lines
between N = N$_{0}$ and N = N$_{0}$ + 2 is positive, and the
vertical distance between isotopic lines with Z = Z$_{0}$ and Z =
Z$_{0}$ + 2 for N $\approx$ N$_{0}$ is larger than for other Z
values.

None of these submagic number effects is shown by the predicted
thin lines systematics. (See also refs. \cite{lmz00,liz76}.)

In the SSME non-smooth abrupt local changes, which are associated
with subshell and deformation effects, are assumed to have been
smoothed out by configuration interaction (eq.~(4) of ref.
\cite{lmz00}), and the mass equation describes a smooth surface
representing their average. The deviations from the average are
mostly small, though, resulting in the above mentioned overall
high-quality predictive power. For the 57 experimental
$Q_{\alpha}$ values measured in region B after the equation was
adjusted \cite{lir73} the respective average deviation
$(\delta_{av})$ and rms deviation $(\delta _{rms})$ of the
predicted values from the data are $-$8 and 220 keV (table II of
ref. \cite{lmz00}). (For all the 115 data points shown in fig. 1,
both experimental and estimated from systematics, the
corresponding values are $-$33 and 214 keV.)

\section{Fusion-evaporation seemingly-well-identified $\alpha$-decay
chains of elements 106-112}

We now address the agreement of the equation with $\alpha$-decay
energies observed in the cold fusion production experiments of
elements 107-112
\cite{mun81,mun84,mun82,mun88,hof95a,hof95b,hof96,hof00} (see also
refs. \cite{hom00} and \cite{arm00}).

Figs. 2-5 show the measured and the predicted $Q_{\alpha}$ values,
assuming that the decays go through or near the g.s. Fig. 5 shows
as well the $Q_{\alpha}$ values observed in hot fusion experiments
producing $^{265}$Sg \cite{laz94,tur98} and $^{267}$Bh
\cite{wil00,eic00}\footnote{In the Introduction of ref.
\cite{lmzla} we were unaware of ref. \cite{eic00}, and ref.
\cite{wil00} was considered among those with an unidentified
evaporation residue parent (like in sect. IV). We are indebted to
Dr. A. T\"urler for  clarification of the situation.}, where in
refs. \cite{tur98} and \cite{eic00} the $\alpha$-chains were
studied after chemical separation of the recoiling evaporation
residue (EVR) parent, leading to chemical identification.

The figures show as well the $\delta_{av}$ and $\delta_{rms}$
values for the plotted data points. On the whole they are not
large, with $\delta_{rms}$ values compatible  with the above
quoted 220 keV value from table II of ref. \cite{lmz00}. The
respective overall $\delta_{av}$ and $\delta_{rms}$ values for all
the 32 data points in figs. 2-5 are 86 and 337 keV.

The largest individual deviations occur for Z = 107, 108 and 110
in fig. 5, with respective values of 659, 712 and $-$605 keV.
These seem to be too large for the above quoted value of 220 keV,
and they correspond to a kink observed in the $^{277}$112
$Q_{\alpha}$ chain at Z = 108. If this kink is a genuine submagic
number effect due to the crossing of N = 162 on the way from Z =
108 to Z = 110 \cite{hof96,hom00}, then fig. 5 illustrates the
oversmoothness of the predicted masses as compared to the
empirical data with its discontinuities in the vicinity of doubly
submagic nuclei. (See sect. II). The pertinent submagic numbers in
the present case are N$_{0}$ = 162 and Z$_{0}$ = 108.

Closer scrutiny of figs. 1 and 5 reveals stronger magicity of
$^{270}$Hs as compared to $^{252}$Fm.

\section{The interpretation problem for some recent hot fusion
experiments with not-well-identified $\alpha$-decay chains}

In hot fusion-evaporation SHE experiments, when the compound
nucleus (CN) is  produced at a sufficiently high excitation energy
(E$_{x}^{cn}$) ,  there are often several conceivable open
emission channels for forming the parent, which would lead to
different observable $\alpha$-decay chains. When the
$\alpha$-decay chain starting from the EVR parent nucleus does not
connect to a known daughter isotope, comparing different
predictions to the observed data might facilitate the choice
between different conceivable interpretations. We consider from
this point of view some recent SHE experiments
\cite{oga01a,oga00,oga01b,oga99}.\footnote{We do not consider here
the experiment described in ref. \cite{nin99} which has not yet
been confirmed. It has been considered from the present point of
view in the Introduction of ref. \cite{lmzla}.}

In refs. \cite{oga01a,oga01b} a $^{248}$Cm target was bombarded by 240 MeV
$^{48}$Ca ions. Three observed three-members $\alpha$-decay chains are
assigned to the nuclide $^{292}$116 and its sequential decay down to
$^{280}$110.

The CN formed in the reaction is $^{296}$116 at E$_{x}^{cn}
\approx$ 27 MeV obtained from the Cm and Ca masses \cite{auw95}
and the predicted mass of the CN \cite{lmz00,lmzla}. At this
higher energy more channels for particle emission might be open
than in the cold fusion experiments considered in sect. III,
including up to $3n$ and also $p$  or $\alpha$ emission.  Four
conceivable EVR parents in addition to the assigned parent
$^{292}$116 are considered in table I.  Their corresponding
formation channels, the estimated values of their excitation
energies (E$_{x}^{evr}$)\footnote{The estimated values of
E$_{x}^{evr}$ in tables I and II are obtained from the kinematics
of the reactions assuming that the evaporated neutrons have zero
kinetic energy and the evaporated charged particles ($p$ and
$\alpha$) have a kinetic energy which is equal to their potential
energy at the top of the Coulomb barrier. Higher kinetic energies
of the evaporated particles would reduce the estimates given in
the tables. This is the case for the $xn$ channels, with an
average C.M. neutron kinetic energy of about 1 MeV per neutron
\cite{nix69}.} when formed, and the deviations $\delta_{av}$ and
$\delta_{rms}$ from the data of the corresponding predicted
$Q_{\alpha}$ values, assuming that the $\alpha$-decays go through
or near the g.s., are given in the table.

For the $4n$ formation channel assigned by the authors the
estimated E$_{x}^{evr}$ value of the parent nucleus $^{292}116$ is
too low (less than 2 MeV) to have a reasonable production cross
section. Moreover, the 758 keV $\delta_{rms}$ value of the
predicted $Q_{\alpha}$ values is too large. On the other hand, for
the $2n$ channel leading to the EVR parent $^{294}116$ the
corresponding estimated E$_{x}^{evr}$ value and the value of
$\delta_{rms}$ are $\leq$14 MeV and 466 keV which are reasonable.
This might lend support to a scenario based on the $2n$ (and
possibly also $3n$) channel.

The last two members of the $\alpha$-decay chains seen in this
experiment agree with the two-members $\alpha$-decay chains
observed before in a Z = 114 experiment \cite{oga00}. If they are
the same, a formation of the present Z = 116 parent by $2n$ (or
$3n$) emission would imply the same formation channel for the Z =
114 EVR parent in ref. \cite{oga00}, rather than the assigned $4n$
channel.

Fig. 6 compares the experimental $Q_{\alpha}$ values with the
predictions of eq. (1) for assumed $2n$, $3n$ and $4n$ evaporation
channels. The improvement when going from $4n$ to $2n$ is clear.

Likewise, fig. 10 of ref. \cite{oga01c} shows the improved
agreement with macroscopic-microscopic $Q_{\alpha}$ systematics
\cite{smo97} achieved by adopting a $2n$ assignment as compared to
$4n$.

As a second example we consider the experiment reported in ref.
\cite{oga99}. A $^{244}$Pu target was bombarded by 236 MeV $^{48}$Ca ions.
An observed three-members $\alpha$-decay chain is considered a good candidate
for originating from the parent $^{289}$114 and its sequential decay down to
$^{277}$Hs (Z = 108).

The CN formed in the reaction is $^{292}$114 at E$_{x}^{cn}
\approx$ 27 MeV \cite{lmz00,lmzla,auw95}. Details for four
conceivable EVR parents are given in table II.

For the $3n$ formation channel assigned by the authors, leading to
the EVR parent $^{289}114$, the respective estimated E$_{x}^{evr}$
value and the $\delta_{rms}$ value are $\leq$9 MeV and 905 keV.
The $\delta_{rms}$ value is too large. If the assignment of the
authors is confirmed this might indicate that the decay chain does
not proceed through levels in the vicinity of the g.s. On the
other hand, for the $p$ or $\alpha$ channels, leading to the
respective EVR parents $^{291}$113 or $^{288}$112,  the respective
E$_{x}^{evr}$ and $\delta_{rms}$ values are 8 or 9 MeV and 414 or
363 keV. This might lend some support to scenarios based on the
$p$ or $\alpha$ evaporation channels.

Fig. 7 compares the experimental $Q_{\alpha}$ values with the
predictions of eq. (1) for assumed $2n, 3n, p$ and $\alpha$
evaporation channels. The advantage of the $p$ or $\alpha$
assignments over $3n$ and $2n$ is obvious.

\bigskip

We thank Yuri Lobanov and Yuri Oganessian for prepublication results
of refs. \cite{oga01a,oga00,oga01b} and Andreas T\"urler for ref.
\cite{eic00}.

\newpage
\begin{table}
    \caption{Conceivable EVR parents of the $\alpha$-decay chain
    [25,27] with their formation channels, their estimated
    values of E$_{x}^{evr}$,  and the deviations $\delta_{av}$ and
    $\delta_{rms}$ from the data of the corresponding predicted
    $Q_{\alpha}$ values.}
    \begin{tabular}{lcccc}
%\hline
        EVR  & Evaporation  & Estimated$^{a}$ E$_{x}^{evr}$
         & $\delta_{av}$ & $\delta_{rms}$  \\
        parent  & channel & (MeV) & (MeV) & (MeV)  \\
        \hline
        $^{294}$116 & $2n$ & 14 & 0.177 & 0.466 \\
        $^{293}$116 & $3n$ & 7 & 0.423 & 0.577  \\
        $^{292}$116 & $4n$ & 2 & 0.669 & 0.758  \\
        $^{295}$115 & $p$ & 7 & -0.767 & 0.905  \\
        $^{292}$114 & $\alpha$ & 8 & -0.590 & 0.699  \\
        \hline
        $^{a}$See footnote 4.&&&&\\
    \end{tabular}
    \label{tableI}
\end{table}

\begin{table}
    \caption{Conceivable EVR parents of the $\alpha$-decay chain
    [28] with their formation channels, their estimated  values
    of E$_{x}^{evr}$, and the deviations $\delta_{av}$ and
$\delta_{rms}$ from the
    data of the corresponding predicted $Q_{\alpha}$ values.}
    \begin{tabular}{lcccc}
%    \hline
        EVR  & Evaporation  & Estimated$^{a}$  E$_{x}^{evr}$
         & $\delta_{av}$  & $\delta_{rms}$  \\
        parent & channel & (MeV) & (MeV) & (MeV) \\
        \hline
        $^{290}$114 & $2n$ & 16 & 0.643 & 0.720  \\
        $^{289}$114 & $3n$ & 9 & 0.847 & 0.905  \\
        $^{291}$113 & $p$ & 8 & -0.241 & 0.414  \\
        $^{288}$112 & $\alpha$ & 9 & -0.181 & 0.363  \\
        \hline
        $^{a}$See footnote 4. &&&& \\
    \end{tabular}
    \label{tableII}
\end{table}

\newpage

\begin{figure}
\begin{center}
\leavevmode \epsfysize=18.0cm \epsfbox{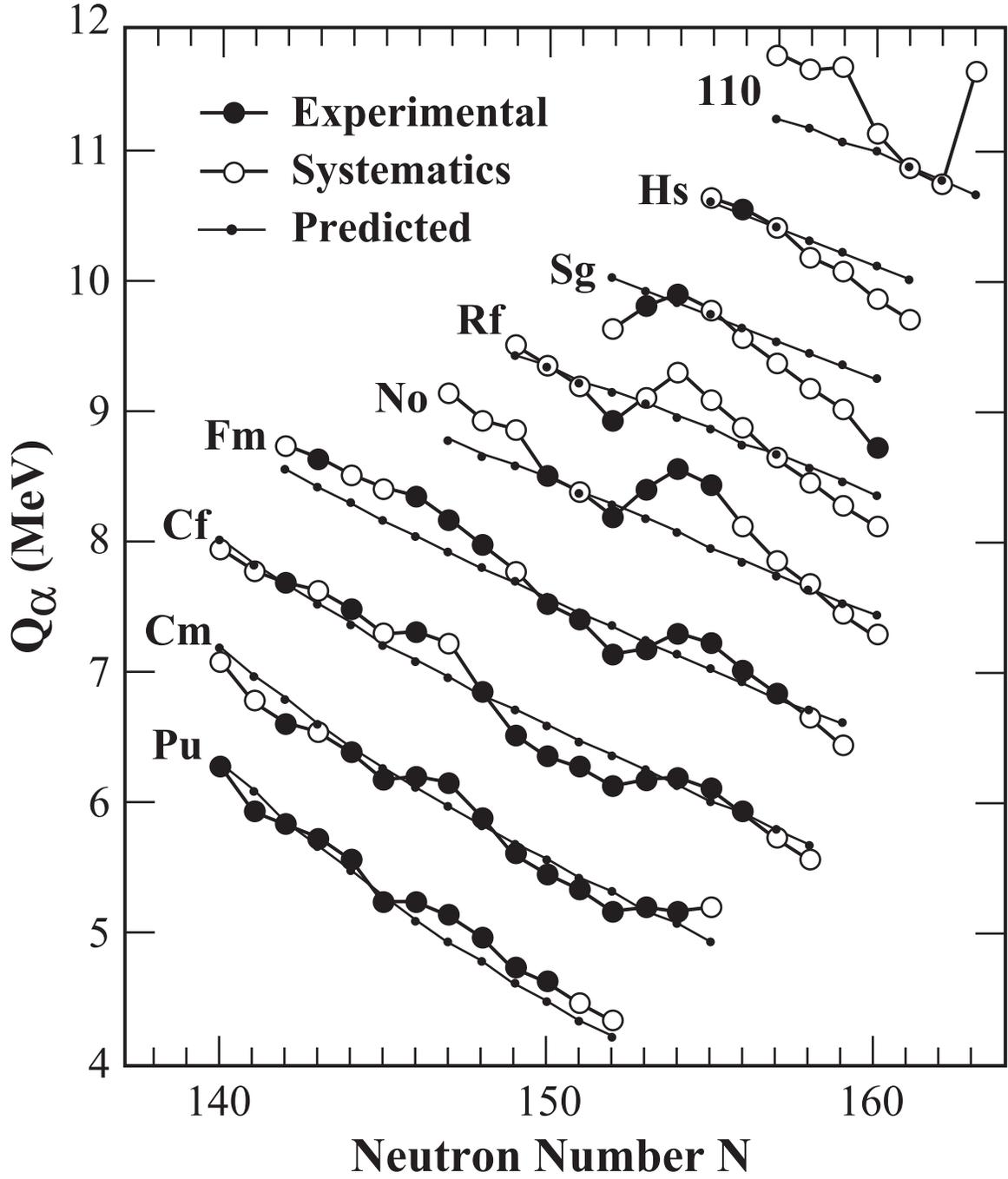}
\end{center}
\caption{Q$_\alpha$ systematics of even-Z elements from Pu through
Z = 110 for N $\geq$ 140. Experimental values and values estimated
from systematics are taken from ref. [6] and predictions from
refs. [1,5].}
\end{figure}

\newpage

\begin{figure}
\begin{center}
\leavevmode \epsfysize=8.0cm \epsfbox{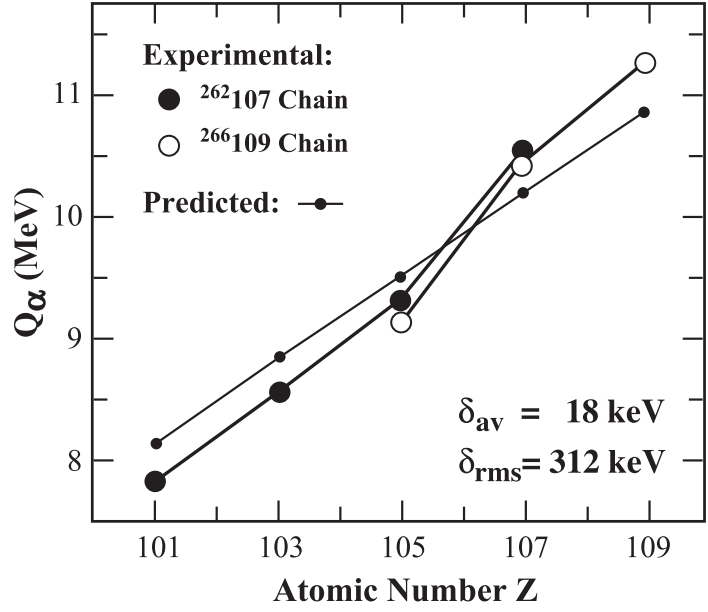}
\end{center}
\caption{Experimental $Q_{\alpha}$ values of the neutron excess $I
= 48$ $\alpha$-decay chains starting from the parent nuclei
$^{262}$107 [11] and $^{266}$109 [13,14]. The predictions are
taken from refs. [1,5].}
\end{figure}

\begin{figure}
\begin{center}
\leavevmode \epsfysize=8.0cm \epsfbox{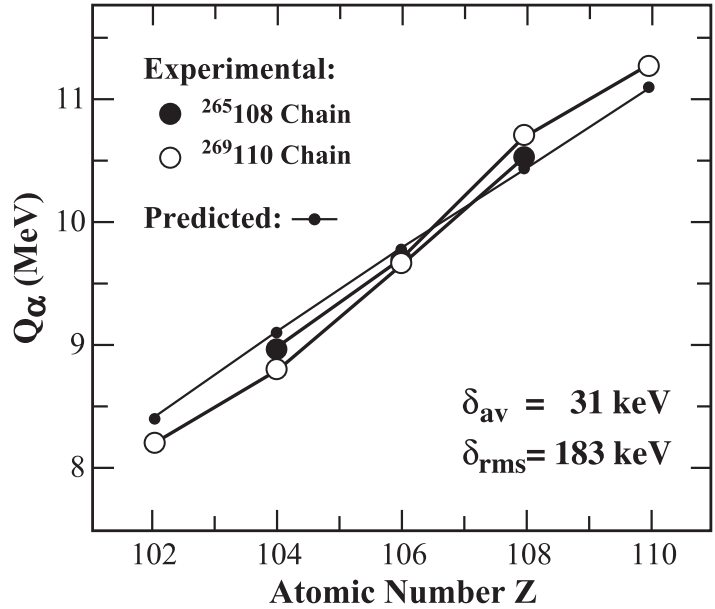}
\end{center}
\caption{Like fig. 2 for the $I = 49$ $\alpha$-decay chains
starting from the parent nuclei $^{265}$108 [12] and $^{269}$110
[15].}
\end{figure}

\newpage

\begin{figure}
\begin{center}
\leavevmode \epsfysize=8.0cm \epsfbox{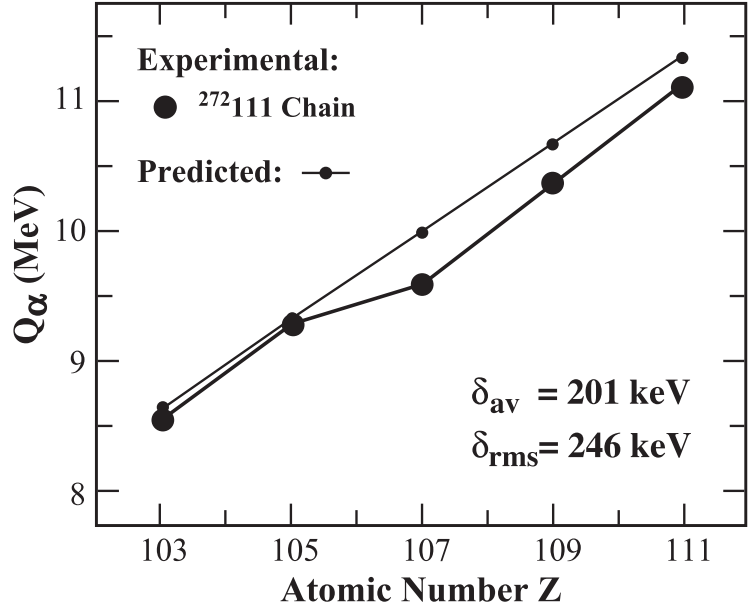}
\end{center}
\caption{Like fig. 2 for the $I = 50$ $\alpha$-decay chain
starting from the parent $^{272}$111 [16,18].}
\end{figure}

\begin{figure}
\begin{center}
\leavevmode \epsfysize=8.0cm \epsfbox{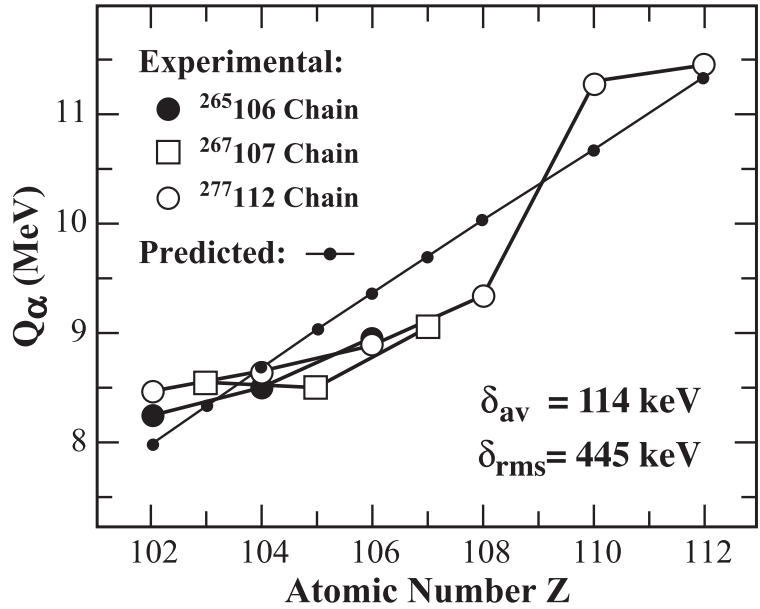}
\end{center}
\caption{Like fig. 2 for the $I = 53$ $\alpha$-decay chains
starting from the deformed g.s. of $^{277}$112 [17-19], from
$^{265}$Sg [21,22] and from $^{267}$Bh [23,24].}
\end{figure}

\newpage

\begin{figure}
\begin{center}
\leavevmode \epsfysize=8.0cm \epsfbox{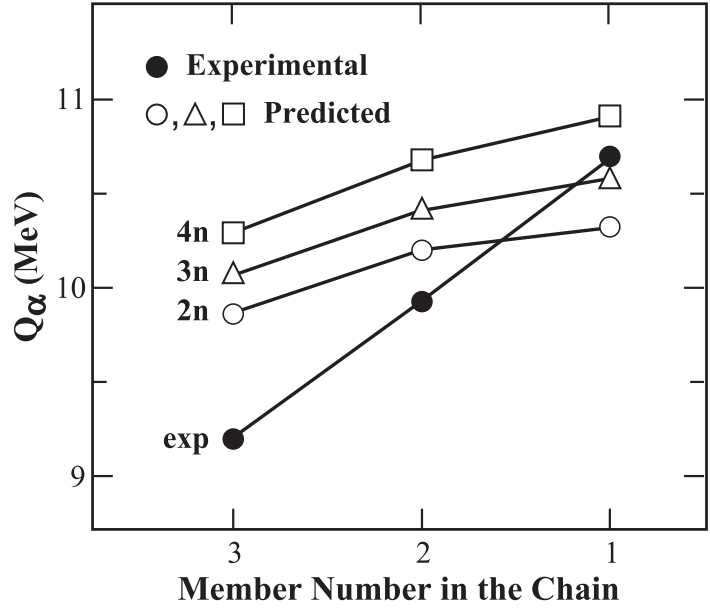}
\end{center}
\caption{Experimental [25,27] (full circles) and predicted [1,5]
$Q_{\alpha}$ values for assumed $2n$ (empty circles), $3n$
(triangles) and $4n$ (squares) evaporation channels in the
$^{48}$Ca on $^{248}$Cm reaction leading to the $^{296}$116 CN.}
\end{figure}

\begin{figure}
\begin{center}
\leavevmode \epsfysize=8.0cm \epsfbox{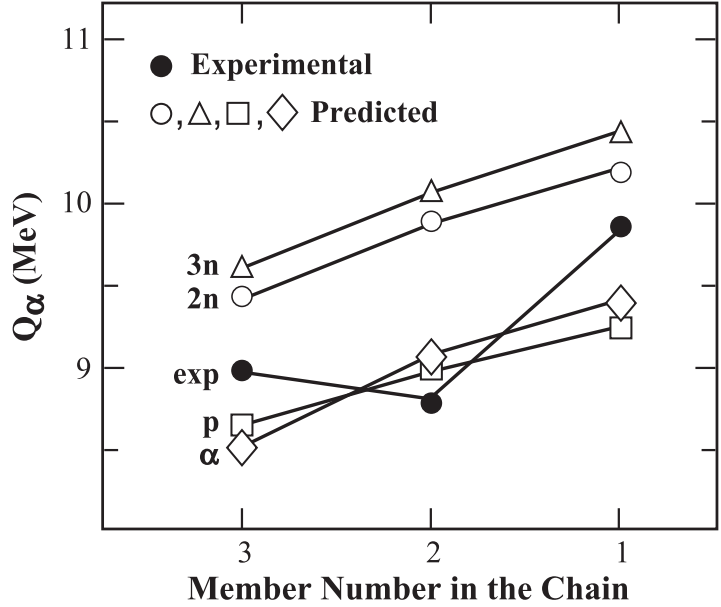}
\end{center}
\caption{Experimental [28] (full circles) and predicted [1,5]
$Q_{\alpha}$ values for assumed $2n$ (empty circles), $3n$
(triangles), $p$ (squares) and $\alpha$ (diamonds) evaporation
channels in the $^{48}$Ca on $^{244}$Pu reaction leading to the
$^{292}$114 CN.}
\end{figure}

\end{document}